\documentclass[twocolumn,showpacs,amsmath,amssymb,superscriptaddress]{revtex4}

\usepackage{epsfig,psfrag}
\usepackage{dcolumn}
\usepackage{bm}
\usepackage{graphicx}
\usepackage{color}
\newcommand{\beq}{\begin{equation}}
\newcommand{\eeq}{\end{equation}}
\newcommand{\beqr}{\begin{eqnarray}}
\newcommand{\eeqr}{\end{eqnarray}}
\newcommand{\e}{{\epsilon}}
\newcommand{\s}{{\sigma}}

\newcommand{\zh}{{\hat{z}}}

\newcommand{\barrho}{{\bar{\rho}}}
\newcommand{\barrop}{\bar{\bar{\rho}}^p({\bf q})}
\newcommand{\barropm}{\bar{\bar{\rho}}^p({-\bf q})}

\newcommand{\barchi}{\bar{\bar{ \chi}}({\bf q})}
\newcommand{\vPi}{{\vec{\Pi}}}

\def\tv{{\tilde{v}}}

\def\bq{{\mathbf q}}

\def\br{{\mathbf r}}

\def\bR{{\mathbf R}}

\newcommand{\etab}{\mbox{\boldmath $\eta $}}

\def\half{{1\over2}}
\def\third{{1\over3}}

\def\eqa{\begin{eqnarray}}
\def\eea{\end{eqnarray}}

\def\ssc{Sol. State. Commun.}
\def\sst{Semicond. Sci. Technol.}

\def\jpcm{{Jour. Phys. Cond. Mat.}}
\begin{document}
\title{Hamiltonian theory of disorder at $\nu=1/3$}
\author{Ganpathy Murthy}
\email{murthy@pa.uky.edu}
\affiliation{Department of Physics and Astronomy, University of Kentucky, Lexington
KY 40506-0055}
\begin{abstract}
The Hamiltonian Theory of the fractional quantum Hall (FQH) regime
provides a simple and tractable approach to calculating gaps,
polarizations, and many other physical quantities. In this paper we
include disorder in our treatment, and show that a simple model with
minimal assumptions produces results consistent with a range of
experiments. In particular, the interplay between disorder and
interactions can result in experimental signatures which mimic those
of spin textures.
\end{abstract}
\maketitle 
Static disorder is crucial to the phenomenology of the
integer quantum Hall effect (IQHE)\cite{vklitzing,perspectives}. The
strong perpendicular magnetic field quantizes the kinetic energy of
the two-dimensional electron gas into Landau levels. While static
disorder broadens each Landau Level (LL), only the state at the center
of the broadened LL is extended at $T=0$, while all other states are
localized\cite{trugman}. This produces the plateau in the Hall
resistance as the chemical potential $\mu$ traverses the localized
states, and the plateau transition when $\mu$ crosses the extended
state. Merely broadening the LLs is not enough to decrease the
transport gap. Because the transport gap depends on {\it extended}
states, one needs to {\it shift} the energies of the extended states
relative to $\mu$ to change the transport gap.

In the fractional quantum Hall effect (FQHE)\cite{fqhe-ex}, the
qualitative role of disorder is to localize the quasiparticles, while
the quantum Hall condensate carries the Hall current. Theoretical gaps
$\Delta_{Th}$ computed by numerical
methods\cite{haldane-rezayi,jain-cf} are much larger than the measured
gaps $\Delta_{Ex}$\cite{dethlefsen}. Empirically one uses the
expression $\Delta_{Ex}=\Delta_{Th}-\Gamma$\cite{dethlefsen}, where
$\Gamma$ is a measure of the disorder broadening and is assumed to be
the same across many fractions. As mentioned above, disorder
broadening by itself cannot account for the reduction of the transport
gap.

In previous work, the effect of dopant disorder on the magnetoroton
minimum at $\third$ has been investigated\cite{magnetorot-disorder}
within the single-mode and self-consistent Born approximations, and
shows a large broadening of the minimum. Also, pointlike disorder at
$\nu=\third$\cite{sheng} has been treated numerically in a finite-size
system, and the transport gap is found to be suppressed. 

It is the purpose of this Letter to use the Hamiltonian
theory\cite{cf-us} of Composite Fermions to develop an approximate
approach which can treat long-range dopant disorder for arbitrary
fractions in the bulk at nonzero temperatures.  We start with Jain's
Composite Fermion (CF) picture\cite{jain-cf}, in which electrons are
bound to $2s$ units of statistical flux to form CF's. The statistical
flux cancels part of the external magnetic flux, leading the CF's at
the principal fractions $\nu=p/(2sp+1)$ to see just the right
effective field to fill $p$ CF-LL's. Thus, the FQHE of electrons is
mapped into the IQHE of CF's\cite{jain-cf}. Following the
Chern-Simons\cite{cs} approaches, R. Shankar and the present author
developed a Hamiltonian theory\cite{cf-us} to describe the dynamics of
CF's in the LLL, which we now briefly describe.

The electronic coordinate and momentum can be separated into a
cyclotron coordinate $\etab_e$ and a guiding center
coordinate $\bR_e$. In the LLL, $\etab_e$ is frozen in its lowest
energy state, and $\bR_e$ has the commutation relations:
$[R_{ex},R_{ey}]=-il^2$ where $l=\sqrt{h/eB}$ is the magnetic
length. The projected interaction controls the dynamics
\beq
H=\int\ {d^2q\over(2\pi)^2)}\ \tv(q) \barrho(\bq)\barrho(-\bq)
\label{electron-H}\eeq
where $\barrho(\bq)=\sum\limits_{j}e^{-i\bq\cdot\bR_{ej}}$ is the
projected density and $\tv(q)=v(q)\exp{-q^2l^2/2}$.  The main problem
is that the coordinates are not complete, and there is no good
starting point for many-body calculations. Our remedy\cite{cf-us} is
to add an auxiliary vector {\it pseudovortex} coordinate (so-called
because it has the commutation relations a true double vortex would
have in the FQHE, but is a fictitious coordinate), with the
commutation relations $[R_{vx},R_{vy}]=il^2/2\nu$ where we have now
specialized to the case of CF's with two fluxes attached. Together,
the two sets of coordinates $\bR_{ej},\ \bR_{vj}$ can be re-expressed
in terms a complete set of coordinates and momenta for the CF (which
we write un-subscripted):
\beqr
\bR_e=&\br-{l^2\over1+c}\zh\times\vPi\\
\bR_v=&\br+{l^2\over c(1+c)}\zh\times\vPi\\
&{[{\Pi_x},{\Pi_y}]}={i}(1-2\nu)/l^2
\label{re-rv-r-Pi}\eeqr
where $c^2=2\nu$ and the last line shows that the CF velocity
operators have commutation relations appropriate to particles seeing a
reduced effective field. This gives a good starting point, the
Hartree-Fock (HF) state of CF's filling $p$ CF-LL's. However, the
price we pay for introducing extra coordinates is that a set of
constraints must be imposed on physical states, ensuring that the
pseudovortex density does not fluctuate:
$\barchi|\psi_{phys}\rangle=\sum\limits_{j}e^{-i\bq\cdot\bR_{vj}}|\psi_{phys}\rangle=0$
A conserving approximation which respects this condition in the sense
of correlation functions can be carried
out\cite{read3,conserving-me}, but a simple shortcut is
available\cite{cf-us}, which displays many of the properties of the CF
at tree level. This is implemented by constructing the {\it preferred
density} $\barrop=\barrho(\bq)-c^2\barchi$ which has the
correct charge and dipole moment of the CF. When used in combination
with the CF-HF state this approximation produces semiquantitative
agreement with experiment for many
quantities\cite{pol-me,shankar-NMR}. In the following, we will use
this approximation to compute excitation energies with disorder. Our
starting point for the clean system is
\beq H=-E_Z S_z+\half\int\ {d^2q\over(2\pi)^2)}\ \tv(q)\barrop\barropm 
\label{starting-H}\eeq
where now we include an implicit sum over spins in the density
operator, and $E_Z$ is the Zeeman energy.  Applying the HF
approximation to this Hamiltonian, one obtains the energies of the
CF-LL's ($\s=\pm1$ is the spin index)
\beq
\e_{n\s}=-{E_Z\s\over2}+ \half\int{d^2q\over(2\pi)^2}\tv(q)\sum\limits_{m}[1-n_F(m\s)]|\barrop_{mn}|^2
\label{CFHF-gaps}
\eeq
where $\barrop_{mn}$ is the matrix element of the preferred
density operator between the CF-LL's $m,\ n$.  The crucial point to
note is that the energy is strongly dependent on the occupations of
the CF-LL's via CF-exchange terms.

Now we turn to disorder. Efros\cite{efros} pointed out that since a
quantum Hall state is incompressible, it cannot screen the disorder
from the distant dopants linearly. The 2DEG forms compressible puddles
of size $s$ (the distance between the dopants and the 2DEG), with
incompressible strips separating them. The electrons feel a
self-consistent short-range potential which has the natural scale of
$E_c=e^2/\varepsilon l\approx \sqrt{B}$. Contrast this to the
compressible $\nu=\half$ system\cite{cs,read3}, where a
$B$-independent disorder width gives good
agreement\cite{half-disorder-us} with nuclear relaxation rate
data\cite{tracy}.

Inspired by the Efros picture\cite{efros}, we construct a
phenomenological model which treats disorder in the self-consistent
Born approximation and interactions in the HF approximation. Since
disorder can mix the different CF-LL's we allow independent
dimensionless coupling coefficients $\alpha_{mn}$ (the average of the
square of the matrix elements of the disorder potential (in units of
$E_c$) between CF-LL's $m$ and $n$). We obtain:
\beqr
\Sigma_{n\s}(\omega)=&E_c^2\sum\limits_{m} \alpha_{mn} G_{m\s}(\omega)\nonumber\\
G_{ms}(\omega)=&{1\over\omega-\e_{m\s}-\Sigma_{m\s}(\omega)}\nonumber\\
n_F(m\s)=-&\int\ {d\omega\over\pi} {Im(\Sigma_{m\s}(\omega))\over1+\exp-\beta(\omega-\mu)}
\label{self-consistent-eqns}\eeqr
Eqs. (\ref{CFHF-gaps},\ref{self-consistent-eqns}) are iterated to
self-consistency at $T\ne0$, with the global condition of $\third$
filling being maintained by adjusting $\mu$. 

To obtain the transport gaps and other extended excitations, we will
assume, in analogy with the IQHE\cite{trugman}, that the CF states are
localized except at the ``band center'' of the disorder-broadened
CF-LL. This is plausible, since the motion of a CF now occurs in some
(self-consistent) random potential in a set of CF-LL's. There are two
natural ways to identify the band center: (i) As the CF-HF energy
$\e_{m\s}$, or (ii) As the energy at which the density of states of
the disorder-broadened CF-LL is the highest. The author has verified
that the two choices exhibit no qualitative differences and only very
small quantitative ones, and in what follows we will use choice (i).
At this level the theory does not treat magnetoexcitons, except in the
$q\to\infty$ limit, when they converge to the gaps between the CF-LLs.
Note that the structure of the puddles can be complex, involving
Wigner crystallites, etc, but since the extended states lie in the
incompressible strip, the detailed structure of the puddles is
irrelevant for the extended states, which makes the self-consistent
Born approximation plausible.

We use the Zhang-Das Sarma potential\cite{zds} $v(q)=2\pi E_c e^{-\lambda
q}/(ql)$ with the thickness parameter $\lambda=0.6 l$. This choice
makes our CF-HF gap for the clean system reproduce the numerical gap
of $0.103E_c$\cite{haldane-rezayi,jain-cf} for the pure Coulomb
interaction in the LLL. All results we present below are for this
choice and $T=50mK$.

The gaps with disorder are smaller than those for a clean system,
because they are strongly affected by the occupations of the levels,
which in turn are affected by disorder. The $n=0,\s=\uparrow$ CF-LL,
which was fully occupied in the clean system, is now partially
occupied, as are the rest of the CF's are in higher CF-LL's whose
densities of states overlap the $n=0\uparrow$ CF-LL. At
self-consistency, the energy of the $n=0\uparrow$ CF-LL  {\it
increases} compared to the clean system, while that of all other
CF-LL's {\it decreases}.  We define the transport gap as $2\times\
min(\mu-\e_{0,\uparrow},\e_{0,\downarrow}-\mu,\e_{1,\uparrow}-\mu)$. The
closest extended state to the chemical potential is always the center
of the $0,\uparrow$ CF-LL. Thus, exactly at $\third$, the residual
carriers should be quasiholes.

\begin{figure}[]
\includegraphics[height=3in,width=2.5in]{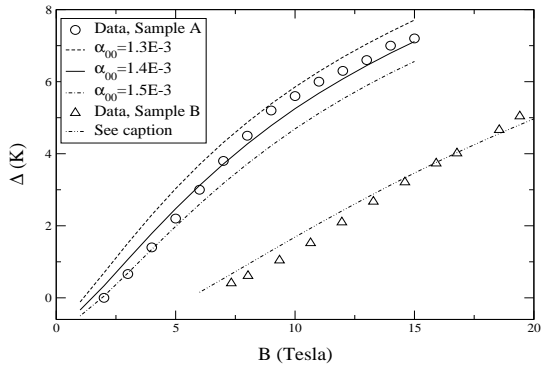}
\caption{A comparison of experimental gaps at $\third$ of Dethlefsen
{\it et al}\cite{dethlefsen} as a function of total perpendicular field and the
predictions of our approach.   For sample A, $\alpha_{mn}=5\times10^{-4}$ and vary $\alpha_{00}$, while for sample B we use 
$\alpha_{00\uparrow}=1.65\times10^{-3},\ \alpha_{00\downarrow}=2.2\times10^{-3},\ \alpha_{mn}=5.25\times10^{-4}$.}
\label{fig1}
\end{figure}

Recently, Dethlefsen {\it et al}\cite{dethlefsen} have studied the
transport gap at $\third$ as a function of perpendicular field for two
different samples. Using the empirical fit
$\Delta_{ex}=\Delta_{th}-\Gamma$ they interpret the measurements for
sample A (mobility $4.5\times10^6 cm^2/Vs$) as showing a crossing of
the $n=0\downarrow$ CF-LL with the $n=1\uparrow$ CF-LL with increasing
$E_Z$, while sample B (mobility $3.5\times10^5 cm^2/Vs$) does not show
this. In Figure \ref{fig1} we show the experimental results and the
results of our approach with a few different sets of parameters, all
of which are assumed to be {\it independent of $B$}.
The agreement is at the few percent level. In our approach the system
is not fully polarized even at the highest field. The $n=0\downarrow$
CF-LL crosses the $n=1,\uparrow$ CF-LL around $4.5\ Tesla$ in sample
A.  At low fields (below $2\ Tesla$ for sample A and below $5\ Tesla$
for sample B) the calculated chemical potential lies beneath the band
center of the lowest CF-LL, showing the absence of the QHE.

Finally, a caveat about the assumed $B$-independence of $\alpha_{mn}$:
As $B$ increases, the self-consistent gap increases, and the width of
the Efros strip\cite{efros} also increases as $\sqrt{\Delta}$, and the
disorder can be expected to decrease. On the other hand, as $B$
increases, the thickness ($\lambda$) of the 2DEG in units of $l_0$
increases (like $B^{1/6}$), which will tend to decrease the
gap. However, in the absence of a microscopic theory I have chosen to
keep $\alpha_{mn}$ and $\lambda$ constant rather than introduce yet
other sources of variability.

Let us now turn to inelastic light scattering
experiments\cite{light-scatt}, which can access excitations invisible
to transport measurements. For example, spin-wave excitations can
clearly be seen in polarized states, as can spin-flip
excitations\cite{light-scatt}, and potentially, magnetoroton
excitations. Recently Groshaus {\it et al}\cite{groshaus} have
reported some intriguing measurements in an extremely clean sample
(mobility $\approx7\times10^6 cm^2/Vs$) in a tilted field. They see
two excitations, one which hardly disperses with $E_Z$ which they
identify as the magnetoroton (``MR'') and another mode which disperses
sharply upwards with $E_Z$, which they consequently identify as a
spin-texture (``ST'') mode with spin 2. The two excitations are
observed to have very different $T$-dependences: The ``MR'' mode
strength decreases with increasing $T$ while that of the ``ST'' mode
increases.

\begin{figure}[]
\includegraphics[height=3in,width=2.5in]{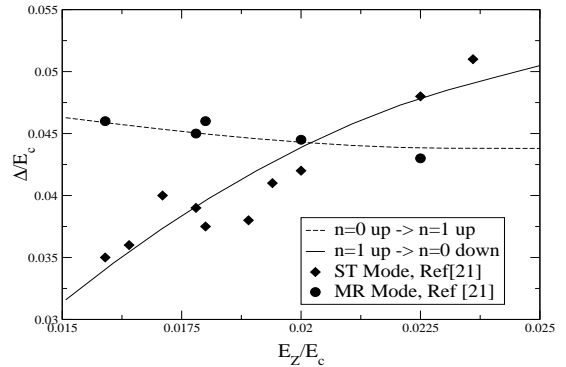}
\caption{A comparison of the light scattering data of Groshaus {\it et
al}\cite{groshaus} to theoretical results for the $n=0\uparrow\Rightarrow
n=1,\uparrow$ and $n=1,\uparrow\Rightarrow n=0,\downarrow$ CF-LL
gaps.}
\label{fig2}
\end{figure}

When comparing our theory to light scattering experiments, we will use
the gaps between CF-LL's (no magnetoexcitons for finite $q$). Within
the Efros picture, the magnetoroton will be strongly influenced by
local environments and considerably
broadened\cite{magnetorot-disorder}, which may lead to a reduced
intensity for light scattering. Since the incident light has a
wavelength $\approx800nm\gg l$ the coherent respose to the light comes
from regions much bigger than the puddles. Purely {\it local}
excitations will span a large range of energies due to differing local
environments. However, the extended states occur at the same energy at
every location in the sample. Thus, it is plausible that transitions
between extended states contribute strongly to light scattering in the
presence of Efros picture disorder.  Figure
\ref{fig2} compares the predicted $n=0\uparrow\Rightarrow
n=1,\uparrow$ and $n=1,\uparrow\Rightarrow n=0,\downarrow$ CF-LL gaps
to the ``MR'' and ``ST'' modes respectively.

Our interpretation of the ``ST'' and ``MR'' modes makes it natural
that, as observed\cite{groshaus}, the strength of the ``ST'' mode
should increase as $T$ increases, while that of the ``MR'' mode should
decrease, due to factors of $n_F(a)(1-n_F(b))$ in the transition
amplitude. As $T$ increases, the occupation of the $n=0,\uparrow$
CF-LL falls and that of the $n=1,\uparrow$ CF-LL increases, leading
to the strengthening of the ``ST'' mode and the weakening of the
``MR'' mode.  The slight reduction in energy of the ``MR'' mode seen
in the data\cite{groshaus} is also reproduced theoretically, and is a
CF-exchange effect.

In summary, we have presented an approach which allows the
phenomenological treatment of disorder in the bulk FQHE at any
$T$. The key ingredients are the Hamiltonian theory\cite{cf-us} of
Composite Fermions\cite{jain-cf}, and the disorder-averaged
self-consistent Born approximation for treating the disorder, modelled
by the Efros picture\cite{efros} of puddles of size $s$ (the distance
between the 2DEG and the dopant layer) separated by incompressible
strips of typical size a few magnetic lengths.

We find that the observed reduction of the gap at $\third$ is the
result of an interplay between the disorder broadening of the CF-LL's
and the strong occupation dependence of the CF-LL energies
(CF-exchange).  An important result is that the interplay of disorder
and interactions produces strong Zeeman dependences of transport and
other gaps which can easily be mistaken for skyrmions\cite{skyrmion}
or other spin structures. As in previous work by the present author at
$\nu=1$\cite{nu1-me}, the observed large slope of the transport
gap\cite{skyrmion-transport} is consistent with an exchange-enhanced
disorder effect. This explanation is in fact not
new\cite{exch-disorder-old}, and is worthy of reexamination in the
light of recent nuclear magnetic resonance (NMR) results, which
indicate that skyrmions are actually localized\cite{skyrmion-loc} at
and around $\nu=1$ at $T=0$. It is also consistent with numerical
work\cite{sinova} showing that localized spin-textures occur in the
ground state for realistic disorder strength. Finally, the view
presented here is consistent with NMR data around $\nu=\third$, which
shows no signatures of skyrmions\cite{pol-third}. To constrain the
theory, it would be desirable to have transport gaps, inelastic light
scattering gaps, polarizations, etc.  for the same sample.

A key assumption here is that the CF extended state is at the
disorder-averaged HF energy of that CF-LL. While it is plausible that
there is one and only one extended state per CF-LL at $T=0$, there is
no microscopic understanding of where it should lie.  Secondly, in
analysing the transport data\cite{dethlefsen} I have assumed a
constant disorder strength (in units of $e^2/\varepsilon l$). In
reality, the Efros strips will broaden slightly as $B$ increases,
which requires a more sophisticated microscopic theory.  Another whole
class of quite mysterious data on the
compressibility\cite{compressibility} of the FQH states exists, to
which the methods developed here are applicable.

The author is grateful to Annelene Dethlefsen and Rolf Haug for
sharing their data and to Jim Eisenstein, Herb Fertig, Javier
Groshaus, Aron Pinczuk, and Misha Reznikov for illuminating
conversations, and finally, to Misha Reznikov for sending me copies of
the Masters Theses of two of his students. I would also like to thank
the Aspen Center for Physics where this work was conceived and partly
carried out, and the NSF for partial support under DMR-0703992.

\end{document}